

\font\bigbrm=cmbx12

\def\m{m_{\rm Pl}}
\def\t{t_{\rm Pl}}
\def\simls{\hbox{$\,$\raise.5ex\hbox{$<$}
         \kern-1.1em \lower.5ex\hbox{$\sim$}$\,$}}
\def\simgt{\hbox{$\,$\raise.5ex\hbox{$>$}
         \kern-1.1em \lower.5ex\hbox{$\sim$}$\,$}}
\def\evol{{1\over\alpha}(\partial_t-\beta^k\partial_k)}
\def\di{\partial_i}
\def\dj{\partial_j}
\def\dk{\partial_k}
\def\dl{\partial_l}
\def\h{\sqrt{h}}
\def\dt{\partial_t}
\def\half{{\textstyle{1\over2}}}

\def\r#1{[#1]}
\def\Guth{1}
\def\Olive{2}
\def\Smoot{3}
\def\KCMW{4}
\def\HPV{5}
\def\GolPir{6}
\def\GolPirK{7}
\def\LaKuMaI{8}
\def\MTW{9}
\def\Linde{10}
\def\Bra{11}
\def\LaKuMaK{12}
\def\LaKuMaP{13}
\def\Cook{14}
\def\Smarr{15}
\def\GolPirR{16}

  \magnification = \magstep1
  \hsize=6.5truein
  \vsize=9.1truein


{\nopagenumbers
 \parindent=0pt
 \leftskip=20pt

\line{\ }
\vfil
{\bigbrm\centerline{INHOMOGENEOUS INFLATION: NUMERICAL EVOLUTION}}

\vfil
\centerline{{\bf Hannu Kurki-Suonio}
 \footnote*{hkurkisu@pcu.helsinki.fi}
}
\smallskip
{\it\centerline{Department of Theoretical Physics}
    \centerline{P.O. Box 9 (Siltavuorenpenger 20 C)}
    \centerline{SF-00014 University of Helsinki, Finland}
}
{\it\centerline{and}}
{\it\centerline{University of California}
    \centerline{Lawrence Livermore National Laboratory}
    \centerline{Livermore, California 94550}
}
\bigskip
\centerline{\bf Pablo Laguna}
\smallskip
{\it\centerline{Department of Astronomy \& Astrophysics}
    \centerline{The Pennsylvania State University}
    \centerline{University Park, Pennsylvania 16802}
}
\bigskip
\centerline{\bf Richard A. Matzner}
\smallskip
{\it\centerline{Center for Relativity}
    \centerline{The University of Texas at Austin}
    \centerline{Austin, Texas 78712}
}
\vfil
\centerline{\bf Abstract}
\medskip
\parindent=20pt

We describe our 3-dimensional numerical relativity
code for the evolution of inhomogeneous cosmologies.
During the evolution the constraint equations are monitored but not
solved.  The code has been tested against perturbation theory with
good results.  We present some runs of inhomogeneous inflation
with strongly inhomogeneous initial data.

\vfil\eject
}


\bigskip
\centerline{\bf I. INTRODUCTION}
\medskip

Inflation\r{\Guth,\Olive} has become the favorite explanation for both
the large-scale homogeneity and the small-scale inhomogeneity of the
universe.  The recent observation of the cosmic microwave anisotropy by
the COBE satellite\r\Smoot\ is consistent with the inflationary universe
scenario.  Most work on inflation has been in the context of either a
flat background spacetime or perturbation theory.
A consistent general-relativistic treatment of strongly
inhomogeneous inflation appears to require numerical relativity.

Previous numerical relativity calculations of inhomogeneous inflation
have been 1-dimensional\r{\KCMW--\GolPirK}.
We have written a 3-dimensional numerical relativity code for this
purpose.  The code consists of two parts.  The first part solves
the initial value problem.  We described it in Ref.~\r\LaKuMaI,
where the motivation and background of this project are also discussed.

Here we present the second part, the evolution code. The main focus of
this paper is on the code rather than physical conclusions about
inhomogeneous inflation.  Thus we discuss the evolution equations
(Sec.~II), numerical techniques (Sec.~III), and code tests (Sec.~IV).
In Sec.~V we present a couple of strongly inhomogeneous
inflation runs.

\bigbreak
\centerline{\bf II. EVOLUTION EQUATIONS}
\medskip

We derive evolution equations for general relativity with scalar field
dynamics in a form suitable for numerical solution.
They will be in the 3+1 formalism\r\MTW\ and for a general gauge.

The matter source is a scalar field.  The covariant form of the field
equation is
$$
   g^{\mu\nu}\phi_{;\mu\nu} = V'(\phi),  \eqno(2.1)
$$
where $V(\phi)$ is a potential function, the form of which
we are not restricting. The prime denotes derivative with respect
to the argument. Also, the semicolon denotes covariant derivative with
respect to the 4-metric $g_{\mu\nu} = h_{\mu\nu}-n_\mu n_\nu$, and
$D_\mu$ is the covariant derivative with respect to the 3-metric
 $h_{\mu\nu}$; $n^\mu$ is the unit normal vector to the 3-space slice,
$n^\mu n_\mu = -1$, and $n^\mu h_{\mu\nu} = 0$ (Greek letter indices range
and sum over 0-3; latin letter indices over 1-3).  Since
$$
   h^{\mu\nu}\phi_{;\mu\nu} = D_\mu D^\mu\phi
    + Kn^\mu \partial_\mu\phi  \eqno(2.2)
$$
and
$$\eqalign{
   n^\mu n^\nu\phi_{;\mu\nu} &= n^\nu\partial_\nu(n^\mu\partial_\mu\phi)
                               -n^\nu n^\mu_{;\nu}\partial_\mu\phi \cr
   &= n^\nu\partial_\nu\eta + n^\nu\partial_\mu\phi(K^\mu_\nu+a^\mu n_\nu) \cr
   &= n^\nu\partial_\nu\eta - {1\over\alpha}h^{\mu\nu}\partial_\mu\alpha
      \partial_\nu\phi, \cr } \eqno(2.3)
$$
Eq.~(2.1) becomes
$$
   D_\mu D^\mu\phi + K\eta - n^\nu\partial_\nu\eta
   + {1\over\alpha}h^{\mu\nu}\partial_\mu\alpha\partial_\nu\phi
    = V'(\phi).  \eqno(2.4)
$$
Here we have defined a new variable
$$
   \eta \equiv n^\mu\partial_\mu\phi      \eqno(2.5)
$$
to replace a second-order equation (2.1) with two equations
(2.4,5) first order in time.

Hereafter we use 3+1 coordinates $(t,x^1,x^2,x^3) = (t,x,y,z)$,
with the metric\r\MTW
$$
   ds^2 = -\alpha^2dt^2 + h_{ij}(dx^i+\beta^idt)(dx^j+\beta^jdt). \eqno(2.6)
$$
The scalar field equations just derived are now
$$\eqalignno{
   \evol\eta &= {1\over\h}\di(\h h^{ij}\dj\phi) + K\eta
              +{1\over\alpha}h^{ij}\di\alpha\dj\phi - V'(\phi) &(2.7)\cr
\noalign{\hbox{and}}
   \evol\phi &= \eta. &(2.8)\cr}
$$

The 3+1 form of the Einstein equations is
$$\eqalignno{
   R + K^2 - K_{ij}K^{ij} &= 16\pi G\rho_H &(2.9)\cr
   D_i K^i_j - D_j K      &= 8\pi GS_j     &(2.10)\cr}
$$
$$\eqalignno{
   \partial_t K^i_j &= -D^i D_j\alpha + \alpha\bigl[R^i_j+KK^i_j
      -8\pi GS^i_j +4\pi G\delta^i_j(S-\rho_H)\bigr]\cr
      &\quad + \beta^k\dk K^i_j - K^k_j\dk\beta^i + \dj\beta^k K^i_k &(2.11)\cr
   \partial_t h_{ij} &= -2\alpha K_{ij} + D_i\beta_j + D_j\beta_i. &(2.12)\cr}
$$
Here $R$ and $R^i_j$ are the Ricci scalar and tensor of the 3-metric,
and $G$ is Newton's constant.
Eqs.~(2.9) and (2.10) are the Hamiltonian and momentum constraints.  We do not
concern ourselves with the solution of the constraint equations in this paper,
as we have discussed it elsewhere\r\LaKuMaI.

For a scalar field, the source terms in Eqs.~(2.9--12) will be
$$\eqalignno{
   \rho_H &= \half h^{ij}\di\phi\dj\phi
              +\half\eta^2+V(\phi) &(2.13)\cr
   S^j    &= -\eta h^{ij}\dj\phi &(2.14)\cr
   S^{ij} &= h^{ij}\bigl[
              -\half h^{kl}\dk\phi\dl\phi+\half \eta^2-V(\phi)\bigr]
              +h^{ik}h^{jl}\dk\phi\dl\phi &(2.15)\cr
   S \equiv S^i_i &= {\textstyle-{1\over2}h^{ij}
                      \di\phi\dj\phi
                      +{3\over2}\eta^2-3V(\phi)}. &(2.16)\cr}
$$
The evolution equation for the extrinsic curvature, Eq.~(2.11), becomes
$$\eqalign{
   \evol K^i_j &= KK^i_j - {1\over\alpha}h^{ik}(\dj\dk\alpha-\Gamma^l_{jk}
                  \dl\alpha) + R^i_j \cr
               &\quad - 8\pi G\bigl[h^{ik}\dk\phi\dj\phi
                  +\delta^i_j V(\phi)\bigr]
                  +{1\over\alpha}(K^i_k\dj\beta^k-K^k_j\dk\beta^i).\cr}
   \eqno(2.17)
$$
We shall evolve the trace $K \equiv K^i_j$ and the traceless part
$A^i_j \equiv K^i_j - {1\over3}\delta^i_j K$ separately.  The evolution
equation for the trace is
$$\eqalign{
   \evol K &= K^2 - {1\over\alpha\h}\di\bigl(\h h^{ij}\dj\alpha\bigr) + R\cr
        &\quad - 8\pi G h^{ij}\di\phi\dj\phi - 24\pi GV(\phi).\cr}
                                                                \eqno(2.18)
$$

Using
$$
   D_i\beta_j = \di\beta_j - \half\beta^l(\dj h_{li}
                + \di h_{lj} - \dl h_{ij}), \eqno(2.19)
$$
the 3-metric evolution equation, Eq.~(2.12), becomes
$$
   \evol h_{ij} = -2K_{ij} + {1\over\alpha}(h_{ki}\dj\beta^k
                                           +h_{kj}\di\beta^k). \eqno (2.20)
$$
It turns out to be useful to have a separate evolution equation for
the square-root of the determinant of the 3-metric, $\h$.  Using
$$
   {1\over\h}\dt\h = \half h^{ij}\dt h_{ij}  \eqno(2.21)
$$
with Eq.~(2.12) we get
$$
   \evol\h = -\h K + {1\over\alpha}\h\dk\beta^k.  \eqno(2.22)
$$

To have a complete set of evolution equations we need to specify equations for
the lapse $\alpha$ and the shift $\beta^i$, {\it i.e.}, to choose a gauge.
The simplest choice is the synchronous gauge,
$$\eqalign{
   \alpha &\equiv 1\cr
   \beta^i &\equiv 0.\cr} \eqno(2.23)
$$
Another choice particularly straightforward to implement in this context
is the harmonic gauge,
$$\eqalignno{
   \evol\alpha &= -\alpha K &(2.24)\cr
   \evol\beta^i &= -\alpha\biggl[h^{ij}{\dj\alpha\over\alpha}
                   +{1\over\h}\dj(\h h^{ij})\biggr]. &(2.25)\cr}
$$
The evolution equations, Eqs.~(2.7,8,17,18,20,22), are valid for any gauge
and we will retain this property as we code them.  We do not attempt
to simplify them by choosing a particular gauge.

We code these evolution equations in conservative form.  The conservative
form of Eq.~(2.22) is
$$
   \dt\h-\dk(\beta^k\h) = -\alpha\h K.  \eqno(2.26)
$$
{}From this follows a general conversion rule:
If the original form is
$$
   (\dt-\beta^k\dk) f = g,  \eqno(2.27)
$$
then the conservative form is
$$
   \dt\tilde f - \dk(\beta^k\tilde f) = \tilde g - \alpha K\tilde f,
   \eqno(2.28)
$$
where
$$
   \tilde f \equiv \h f,  \eqno(2.29)
$$
{\it etc}.

Our final set of equations is
$$\eqalignno{
   \dt\tilde K-\dk(\beta^k\tilde K) &= -\di(\tilde h^{ij}\dj\alpha) + \alpha\h
R
      - 8\pi G\alpha\tilde h^{ij}\di\phi\dj\phi
      - 24\pi G\alpha\h V(\phi)                           &(2.30)\cr
   \dt\tilde A^i_j-\dk(\beta^k\tilde A^i_j) &=
      - \tilde h^{ik}(\dj\dk\alpha-\Gamma^l_{jk}\dl\alpha)
      + \alpha\tilde h^{ik} R_{kj}
      - 8\pi G\alpha\tilde h^{ik}\dk\phi\dj\phi  \cr
     &\quad - {1\over3}\delta^i_j\bigl[
         - \dk(\tilde h^{kl}\dl\alpha) + \alpha\h R
         - 8\pi G\alpha\tilde h^{kl}\dk\phi\dl\phi \bigr] \cr
     &\quad + \dj\beta^k\tilde A^i_k - \dk\beta^i\tilde A^k_j    &(2.31)\cr
   \dt\tilde\eta-\dk(\beta^k\tilde\eta) &= \di(\alpha\tilde h^{ij}\dj\phi)
      - \alpha\h V'(\phi)                                  &(2.32)\cr
   \dt\h-\dk(\beta^k\h) &= - \alpha\h K                    &(2.33)\cr
   \dt\tilde h_{ij}-\dk(\beta^k\tilde h_{ij}) &=
      - {\textstyle{5\over3}}\alpha K\tilde h_{ij}
      - 2\alpha A^l_j\tilde h_{il}
      + \dj\beta^l\tilde h_{il} + \di\beta^l\tilde h_{lj}  &(2.34)\cr
   \dt\tilde\phi-\dk(\beta^k\tilde\phi) &=
      - \alpha K\tilde\phi + \alpha\tilde\eta.             &(2.35)\cr}
$$

The harmonic gauge equations in conservative form are
$$\eqalignno{
   \dt\tilde\omega-\dk(\beta^k\tilde\omega) &= 0           &(2.36)\cr
   \dt\tilde\beta^i-\dk(\beta^k\tilde\beta^i) &=
      - \alpha K\tilde\beta^i
      - \alpha^2\biggl[\tilde h^{ij}{\dj\alpha\over\alpha}
         + \dj\tilde h^{ij}\biggr],                        &(2.37)\cr}
$$
where $\omega \equiv 1/\alpha$, or $\tilde\omega \equiv \h/\alpha$.

We need initial data that satisfies the constraint equations (2.9) and (2.10),
which we can write as
$$\eqalignno{
   R + {\textstyle{2\over3}}K^2-A^i_jA^j_i
   &= 8\pi G\bigl[h^{ij}\di\phi\dj\phi+\eta^2+2V(\phi)\bigl] &(2.38)\cr
   \di A^i_j+\Gamma^i_{li}A^l_j-\Gamma^l_{ji}A^i_l-{\textstyle{2\over3}}\dj K
   &= -8\pi G\eta\dj\phi.                                     &(2.39)\cr}
$$
We discussed the procedure for solving the initial value problem in
Ref.~\r\LaKuMaI.  In addition to these numerically solved initial data, we have
also used data that satisfies the constraints trivially.

This latter method, previously used by Goldwirth and Piran\r\GolPir, involves
two scalar fields to achieve a momentarily flat space for the initial slice.
One of the scalar fields is the inflaton field, $\phi$, which can be set
arbitrarily according to the problem we wish to solve.  The other field,
$\phi_R$, is an auxiliary field, with $V(\phi_R) \equiv 0$.
We want the right hand side of Eq.~(2.38) to be constant,
and that of Eq.~(2.39) to be zero
on the initial slice.  This is achieved by setting
$$\eqalign{
   \eta &\equiv 0 \cr
   \phi_R &\equiv 0 \cr
   \eta_R &\equiv \bigl[2\rho_H- h^{ij}\di\phi\dj\phi
                  -2V(\phi)\bigr]^{1/2}, \cr}             \eqno(2.40)
$$
where $\rho_H =$ const.
Equations (2.38) and (2.39) can now be satisfied with
$$\eqalign{
   K &\equiv -3\biggl({8\pi G\rho_H\over3}\biggr)^{1/2} \cr
   A^i_j &\equiv 0 \cr
   h_{ij} &\equiv \delta^i_j. \cr}                             \eqno(2.41)
$$
The initial slice will thus have a homogeneous total energy density,
but with an inhomogeneous composition.  The energy density becomes
inhomogeneous immediately when it is evolved.
In a typical run, $\phi_R$ soon redshifts away whereas the vacuum energy
$V(\phi)$ keeps the inflaton field dynamically important.

\bigbreak
\centerline{\bf III. NUMERICAL TECHNIQUES}
\medskip

Replacing the continuous spacetime with a discrete grid requires a decision
about the relative positioning of variables on the grid.
As discussed below, we have divided the variables into
two sets staggered in time.
This division arises naturally from the form of the equations, although some
terms in the equations violate this structure and require interpolation.

In the previous section, Eqs.~(2.30--35) are written
in the order they are actually evolved
in our code.  The lapse is calculated before $\h$, and the shift
before $\tilde K$.  This ordering is determined by what quantities are
needed to evolve which quantity.
The quantities $\alpha$, $\h$, $\tilde h_{ij}$, and $\tilde\phi$
are evaluated after the full time-step, {\it i.e.}, at times
$$
   t^{n-1}, t^n\equiv t^{n-1}+\Delta t^n, \ldots,
$$
where $\Delta t^n$ is the $n^{\rm th}$ time-step.
The quantities $\beta^i$, $\tilde K$, $\tilde A^i_j$, and $\tilde\eta$
are evaluated at half-time, {\it i.e.}, at times
$$\textstyle
   t^{n-1}+{1\over2}\Delta t^n, t^n+{1\over2}\Delta t^{n+1},\ldots.
$$
With the possible exception of lapse and shift, this scheme does not require
any extrapolation in time.  All quantities that are needed to evolve a
particular quantity have already been evolved to the time level at which
they are needed.  The quantities $\alpha$, $\beta^i$, $\h$, and $K$ are
needed at both full- and half-time, so they are interpolated when needed.
Depending on the choice of gauge, the calculation of $\alpha$
and $\beta^i$ at a new time level may require extrapolation of some
quantities in time, and should then be iterated.

It is a common practice to have a similar staggering in space.  A gradient of
a variable would naturally lie between the grid points where the variable is
located.  In a 3-d code this becomes complicated.  Picture the grid dividing
the space into small cubes, or zones.  Some variables might naturally lie at
zone centers, others at zone faces, and yet others at zone edges.  The grid
will have 3 times as many faces and edges as zones (in 3 different
orientations).  Typically the different components of a 3 vector would lie on
differently oriented faces.  If the equations respect such a structure, the
extra coding effort will pay off by giving a good accuracy with a small number
of zones.  We, however, did not find our equations to naturally fit into such
a scheme.  Because so many terms would have to be interpolated, it appeared
unlikely that this would lead to much improved performance.  Therefore we
decided to position all the variables the same way in space.  We refer to
these locations simply as grid points.  To obtain a value of a spatial
derivative at a grid point we thus need to difference over two grid spacings.

For simplicity and generality, we are using Cartesian coordinates.  We have
coded for nonuniform grid spacings in $x$, $y$, and $z$.  However, to maintain
second-order accuracy in space with a nonuniform grid the spacing should not
vary too steeply.  A variable will lie at a gridpoint $(x,y,z)$, where, {\it
e.g.}, $x$ takes on any of the values $x_1, \ldots, x_{N_x}$, and
$$
   x_n = x_{n-1}+\Delta x_n.                                    \eqno(3.1)
$$
The first and last grid points in each direction, {\it e.g.}, at $x_1$ and
$x_{N_x}$, are dummies.  Variables are not evolved at these grid points but
are set after the evolution according to the boundary condition.  Thus the
`physical' grid has only $(N_x-2)(N_y-2)(N_z-2)$ points.

Periodic boundary conditions are
$$
\eqalign{ f(x_{N_x},y,z) &= f(x_2,y,z), \cr
          f(x_1,y,z) &= f(x_{N_x-1},y,z), \cr}                  \eqno(3.2)
$$
{\it etc}.  Reflective boundary conditions for tensors are different than
for scalars, because some components change signs, {\it e.g.},
$$
\eqalign{ \beta^1(x_1,y,z) &= -\beta^1(x_2,y,z), \cr
          h_{12}(x_1,y,z) &= -h_{12}(x_2,y,z). \cr}             \eqno(3.3)
$$

The largest piece of the evolution code is the calculation of the Ricci tensor
$R_{ij}$.  It is calculated from the 3-metric $h_{ij}$ and is needed for
evolving the extrinsic curvature.  In the process we obtain the connection
coefficients $\Gamma^i_{jk}$, also needed for evolving $A^i_j$.
($\Gamma^i_{jk}$ is also needed for checking the momentum constraint, and
$R\equiv R^i_i$ is needed for checking the Hamiltonian constraint).
These are computed directly from their definitions
$$
   \Gamma^i_{jk} \equiv \half h^{il}(\dk h_{lj}+\dj h_{lk}-\dl h_{jk})
                                                                \eqno(3.4)
$$
and
$$ \eqalign{
   R_{ij} &\equiv \dk\Gamma^k_{ij} - \dj\Gamma^k_{ik}
                + \Gamma^k_{lk}\Gamma^l_{ij} - \Gamma^k_{lj}\Gamma^l_{ik} \cr
          &= \half\dk h^{kl}\bigl(\dj h_{li}+\di h_{lj}-\dl h_{ij}\bigr) \cr
          &\quad + \half h^{kl}\bigl(\dk\dj h_{li} + \dk\di h_{lj}
                                    - \dk\dl h_{ij} \bigr) \cr
          &\quad - \half\dj h^{kl}\di h_{lk} - \half h^{kl}\dj\di h_{lk}
              + \Gamma^k_{lk}\Gamma^l_{ij} - \Gamma^k_{lj}\Gamma^l_{ik}.  \cr}
                                                                \eqno(3.5)
$$
(We first invert $h_{ij}$ to obtain $h^{ij}$).
We use the latter form for $R_{ij}$ to avoid taking derivatives of
$\Gamma^i_{jk}$, which already involves derivatives.  With our grid structure
we obtain second derivatives more compactly by calculating them directly.

There are 6 components of $R_{ij}$ and 18 components of $\Gamma^i_{jk}$ to be
calculated, with double sums.  This could have been coded with loops over the
indices, relying on the compiler to unroll the summation loops
to permit vectorizing.  However, there are complications due to the
symmetries in
indices.  Also, many of the terms cancel out  with certain combinations of
indices.  Therefore we have coded in all components separately, with the sums
written out.  We actually wrote a small program  to produce this long
(almost 600 lines of Fortran) but repetitive piece of
code.

The evolution equations are mostly rather straightforward to code.  Since
we are evolving the `twiddled' quantities---see Eq.~(2.29)---which are often
needed `untwiddled', we have to be careful to multiply and divide by $\h$ when
needed.  To make this possible, $\h$ has to be evolved to a new
full-time level first.

The present version of the code does not yet include the
advection terms $\dk(\beta^kf)$,
and is therefore restricted to $\beta^i \equiv 0$.
Also, the right-hand-side terms involving $\beta^i$ have been left out in
the $\tilde A^i_j$ and $\tilde h_{ij}$ equations.  This allows us to evolve
$\tilde A^i_j$ without matrix inversion.

The evolution of $\tilde h_{ij}$ involves matrix inversion since the
right-hand side of Eq.~(2.34)  contains other components of this tensor
in the
$-2\alpha A^l_j\tilde h_{il}$ term.
With $\beta^i\equiv0$, the discretized form of this equation (taking us from
$t^n$ to $t^{n+1}$) can be written as
$$\eqalignno{
   H^{n+1} &= H^n - \half\alpha\Delta t M(H^n+H^{n+1}),              &(3.6)\cr
\noalign{\hbox{or}}
   H^{n+1} &= (1+\half\alpha\Delta t M)^{-1}(1-\half\alpha\Delta t M)H^n,
                                                                &(3.7)\cr}
$$
where $H \equiv (\tilde h_{11}, \tilde h_{12}, \tilde h_{13}, \tilde h_{22},
\tilde h_{23}, \tilde h_{33})$ is a 6-component vector, and $M$ is a
$6\times6$ matrix, whose components are linear combinations of the extrinsic
curvature components.  A matrix inversion at each grid point would be rather
time consuming.  As we only need second-order accuracy in $\Delta t$,
we replace Eq.~(3.7) by
$$
   H^{n+1} = (1-\alpha\Delta t M+\half\alpha^2\Delta t^2M^2)H^n, \eqno(3.8)
$$
which requires mere matrix multiplication, and does not prevent vectorization
over the grid.

A similar procedure will have to be utilized in the evolution of $\tilde
A^i_j$ once nonzero shift is included.  Presently we evolve each component
separately.  Although only 5 of the components are independent, we evolve all
9 of them since reconstructing the remaining components from the minimum set
of 5 is nontrivial.  (Since $\tilde A^i_i = 0$, we could easily skip evolving
one of the diagonal components). The covariant components are symmetric,
so if we rewrote
Eq.~(2.31) to evolve $\tilde A_{ij}$ instead of $\tilde A^i_j$, there would be
only 6 components to evolve.  But this gives a nonlinear equation, whereas
Eq.~(2.31) is linear in $\tilde A^i_j$.

We are presently running free evolution, {\it i.e.}, the constraint equations
are not solved after the initial slice.  To see how accurately the
constraints are maintained, we compare the left- and right-hand sides of
Eqs.~(2.38) and (2.39) during the evolution.

We use three conditions to determine the time step.  The first is the
Courant condition.  The code requires three time steps to carry
information from the grid point $(x_l,y_m,z_n)$ diagonally to $(x_{l+1},
y_{m+1},z_{n+1})$.  The time step must be short enough that this grid
velocity is not less than the speed of light.  To satisfy this condition
everywhere, we need to find the shortest physical distance $\Delta s$,
where
$$
   \Delta s^2 = h_{ij}\Delta x^i\Delta x^j, \eqno(3.9)
$$
between neighboring points. To be more precise, we set
$$
   \Delta t_C = C \min \biggl({\Delta s\over\alpha d}\biggr), \eqno(3.10)
$$
where $d = 1, 2$, or 3 depending on whether the neighboring point
differs in one, two, or three coordinates, and $C<1$.  In practice we
have used $C = 0.7$.

The second condition is an expansion condition.  According to
Eq.~(2.33), the local volume
expansion rate is given by $-\alpha K$.  To keep expansion per time
step below a certain amount $E$, we set
$$
   \Delta t_E = E/\max(\alpha|K|).  \eqno(3.11)
$$
The third condition is a smoothness condition.  To prevent the time step
from increasing too rapidly, we set
$$
   \Delta t_S = (1+S)\Delta t_{\rm previous}.  \eqno(3.12)
$$
The time step is then the smallest of these three,
$$
   \Delta t = \min(\Delta t_C,\Delta t_E,\Delta t_S).  \eqno(3.13)
$$

The Courant condition gives a shorter time step when the number of grid
points is increased.  In the runs presented here, we adjusted the other
two conditions with grid size so that halving the grid spacing also halved
the time step consistently over the whole run.
Thus we have used $E = 0.1(32/N)$ and $S = 0.01(32/N)$, where N is the number
of
grid points in one direction.

In practice, the controlling condition was the Courant condition in the
early part of the run, the expansion condition during most of the
inflation, and the smoothness condition after inflation.
The time-scale of the post-inflation oscillations of $\phi$ is not
related to any of these three conditions.  Indeed, when we ran long
enough, the time step became too long to resolve these oscillations.
We experimented with a fourth condition to control this, but could not
easily find one that is both general and simple.  Instead, we just
chose a small enough value for $S$ to resolve the first
oscillations, as we were not interested in the later ones.
In running different problems it may be useful to experiment with the
time-step control, to optimize it for each case.

\bigbreak
\centerline{\bf IV. CODE TESTS}
\medskip

\bigskip
\centerline{\bf A. Homogeneous tests}
\medskip

The simplest test is a homogeneous one.  We did a homogeneous chaotic
inflation\r\Linde\ run with the potential  $ V(\phi) = {1\over4}\lambda\phi^4$,
$\lambda = 10^{-3}$, and the inflaton field initially at $\phi = 5 \m$.
The perturbation runs below are
perturbations around this homogeneous run.
Figures 1 and 2 show the time evolution of the inflaton $\phi$ and the
scale factor $a$, respectively.
The code deviates from the exact results by less than
$10^{-3}$.  Another run with a different time step showed the error to
be quadratic in time step.  We also ran a Kasner model, where the
expansion rates along the three axes were all different, and obtained
similar accuracy.

\bigbreak
\centerline{\bf B. Perturbation tests}
\medskip

The perturbation theory for inflation has been discussed by
Brandenberger {\it et al.} \r{\Bra}.  The unperturbed spacetime is
assumed to be spatially flat.  We shall work in the synchronous gauge.
The metric and the inflaton field can be written as
$$
ds^2 = - dt^2 + a(t)^2\Bigl\lbrace\bigl[1+A(t,{\bf x})\bigr]\delta_{ij}
       +B(t,{\bf x})_{,ij}\Bigr\rbrace dx^idx^j \eqno(4.1)
$$
and
$$
\phi(t,{\bf x}) = \phi_0(t) + \delta\phi(t,{\bf x}), \eqno(4.2)
$$
where $\delta\phi$, $A$, and $B$ are small perturbations.  The
$0^{\rm th}$ order, or homogeneous, equations are
$$
H^2 \equiv \biggl({\dot a\over a}\biggr)^2 = {8\pi G\over3}\textstyle{
\bigl[{1\over2}\dot\phi_0^2+V(\phi_0)\bigr],} \eqno(4.3)
$$
$$
2{\ddot a\over a} + \biggl({\dot a\over a}\biggr)^2 = -8\pi G\textstyle
{\bigl[{1\over2}\dot\phi_0^2-V(\phi_0)\bigr],} \eqno(4.4)
$$
and
$$
\ddot\phi_0 + 3H\dot\phi_0 + V'(\phi_0) = 0, \eqno(4.5)
$$
where Eq.~(4.3) is the Hamiltonian constraint.  The momentum constraint
vanishes identically to $0^{\rm th}$ order.

We expand the perturbations in plane waves
$$
\delta\phi(t,{\bf x}) = \sum_k\delta\phi_k(t)e^{i{\bf k\cdot
x}}, \eqno(4.6)
$$
{\it etc.}, where ${\bf k}$ is the comoving wave vector.  The first order
perturbation equations for the modes $\delta\phi_k(t)$,
$A_k(t)$, and $B_k(t)$ are then
$$\eqalignno{
3\ddot A_k - k^2\ddot B_k + 3H(3\dot A_k - k^2\dot B_k) + k^2a^{-2}A_k
&= -24\pi G \bigl[\dot\phi_0\delta\dot\phi_k - V'(\phi_0)\delta\phi_k\bigr]
&(4.7)\cr
\ddot A_k + 3H\dot A_k
&= -8\pi G \bigl[\dot\phi_0\delta\dot\phi_k - V'(\phi_0)\delta\phi_k\bigr]
&(4.8)\cr
-k^2a^{-2}A_k - H(3\dot A_k - k^2\dot B_k)
&= -8\pi G \bigl[\dot\phi_0\delta\dot\phi_k + V'(\phi_0)\delta\phi_k\bigr]
&(4.9)\cr
\dot A_k &= -8\pi G\dot\phi_0\delta\phi_k
&(4.10)\cr
\delta\ddot\phi_k + 3H\delta\dot\phi_k + V''(\phi_0)\delta\phi_k
&= \textstyle{-k^2a^{-2}\phi_k - {1\over2}\dot\phi_0(3\dot A_k-k^2\dot
B_k). } &(4.11)\cr}
$$

To obtain `exact' perturbation theory results we solved these ordinary
differential equations with the Runge-Kutta method.  The metric
evolution equations (4.4), (4.7), and (4.8) can here be ignored, and the
metric $(a, A_k, B_k)$ can be evolved using the constraint equations (4.3),
(4.9), and (4.10).

For the perturbation runs with the code we need the relation
between the code variables  $K$ and $A^i_j$, and the perturbation theory
variables $A$ and $B$.  From the metric evolution equation (2.12) we
obtain
$$\eqalignno{
   K_0 &= -3H, &(4.12)\cr
   \delta K& = \textstyle{-{3\over2}\dot A -{1\over2}\nabla^2\dot B,}
&(4.13)\cr
   A^i_j &= \textstyle{-{1\over2}\dot B_{,ij} + {1\over6}\delta_{ij}
      \nabla^2\dot B,} &(4.14)\cr}
$$
with $K = K_0+\delta K$.  The scalar 3-curvature is
$$
   R = -2a^{-2}\nabla^2A. \eqno(4.15)
$$

   The initial data has to satisfy the perturbation theory constraint equations
(4.9) and (4.10).  This is accomplished by choosing arbitrary values for
$\phi_0$, $\dot\phi_0$, $\delta\phi$, $\delta\dot\phi$, $A$, and $B$,
and solving for
$\dot A$ and $\dot B$ from Eqs.~(4.9) and (4.10).  For our test runs
we have chosen initial data of the form
$$\eqalign{
   a &= 1, \cr
   \phi_0 &= 5\m, \cr
   \dot\phi_0 &= 0, \cr
   \delta\phi &= \sum_k\delta\phi_k e^{i{\bf k\cdot x}}, \cr
   \delta\dot\phi &= 0, \cr
   A &= 0, \cr
   B &= 0. \cr} \eqno(4.16)
$$
For this case, the constraint equations (4.9) and (4.10) give
$$\eqalign{
   \dot A_k &= 0, \cr
   \dot B_k &= -{8\pi G\over Hk^2}V'(\phi_0)\delta\phi_k. \cr} \eqno(4.17)
$$
The initial values for the extrinsic curvature variables are thus
$$\eqalign{
   K_0 &= -3H, \cr
   \delta K &= -4\pi G{V'(\phi_0)\over H} \sum_k\delta\phi_k e^{i{\bf k\cdot
      x}}, \cr
   A^i_j &= {4\pi G\over H}V'(\phi_0)\sum_k\biggl( {1\over3}\delta_{ij}
      -{k_ik_j\over k^2}\biggr) \delta\phi_k e^{i{\bf k\cdot x}}, \cr}
   \eqno(4.18)
$$
where
$$
   H = \biggl[{8\pi G\over3}V(\phi_0)\biggr]^{1/2}. \eqno (4.19)
$$

The particular initial data we have used is
a superposition of three perpendicular
plane waves,
$$
   \delta\phi(t_0,{\bf x}) = \delta\phi_1(t_0)e^{ik_1x}
      + \delta\phi_2(t_0)e^{ik_2y}
      + \delta\phi_3(t_0)e^{ik_3z}, \eqno(4.20)
$$
with small and equal amplitudes
$$
   \delta\phi_1(t_0) = \delta\phi_2(t_0) = \delta\phi_3(t_0) = 10^{-6},
   \eqno(4.21)
$$
but different wavelengths $L_i = 2\pi/k_i$,
$$
   L_1 = 0.2H^{-1}, L_2 = 0.3H^{-1}, L_3 = 0.4H^{-1}. \eqno(4.22)
$$
Here the subscripts $i = 1, 2, 3$ denote the three different modes, rather than
components of a vector.  Our grid had one wavelength in each direction.  To get
the same number $N_x = N_y = N_z$ of grid points per each wavelength, we
have used
different grid spacings $\Delta x^i$ in each direction.

We studied the convergence of the code results towards the perturbation theory
results as the number of grid points is increased (the time step and grid
spacing is decreased).  The results are shown in Figs.~3, 4, and 5.  These
show the evolution of the amplitude of the three perpendicular perturbation
waves in $\phi$, $A$, and $B$.

Initially, the waves are well inside the horizon.
(By `horizon' we refer to the
Hubble length, as is common among astroparticle physicists).
The Hubble constant at the initial time is
$$
   H = \biggl({8\pi\lambda\over3\m^2}\biggr)^{1/2}
      \biggl({\phi_0\over\m}\biggr)^2 = 2.29\m. \eqno(4.23)
$$
The initial values of $k_i/H$
were $10\pi$, $20\pi/3$, and $5\pi$.  We have arbitrarily set the initial
time as $t_0 = \t    $.
As the universe inflates, the waves exit the horizon one by one.  Defining the
time of exit by $k_i = H$, this happens at $t = 2.20\t    $, $2.33\t$, and
$2.51\t    $.

The behavior of the perturbations change markedly as they exit the horizon.
Inside the horizon, the initial $\phi$ perturbation oscillates and these
oscillations are damped by the expansion, all three wavelengths at the same
rate.  There was no initial perturbation $A$, $B$,  in the metric,
but these are now induced by the $\phi$ perturbation.  The $A$ perturbation
oscillates with a growing amplitude, a quarter phase behind $\delta\phi$.
The $B$
perturbation exhibits monotonic growth with a small oscillation superimposed.
The metric perturbations don't have time to reach the $\phi$ perturbation
amplitude, before the perturbations exit the horizon.

Outside the horizon the oscillations cease.  The $\phi$ perturbation decreases
while $A$ grows and $B$ stays constant.  At the end of inflation the
growth in $A$ ceases, whereas $\delta\phi$ shows anharmonic oscillation
with $\phi_0$ at the bottom of $V(\phi)$.

The discretization error exhibits a quadratic behaviour when we increase the
number of grid points from $16^3$ to $48^3$.  The errors in $B$ are small,
as are the errors in $A$ and $\delta\phi$ while inside horizon.  The larger
errors in $\delta\phi$ and $A$ outside the horizon are due to the switch from
oscillating to monotonic behaviour, when a small phase error leads to a
relatively large amplitude error.  Even then, the errors
in the final values in the $48^3$ runs
are less than $3\%$, after a volume expansion by
more than $10^{100}$.

\bigbreak
\centerline{\bf C. Memory and time requirements}
\medskip

The code has 56 variables per grid point.  These are $\h$, $h_{ij}$ (6),
$h^{ij}$ (6), $\Gamma^i_{jk}$ (18), $R_{ij}$ (6), $R$, $\alpha$,
$\beta^i$ (3), $K$, $A^i_j$ (9), $\phi$, $\eta$, $\phi_R$, and $\eta_R$.
In addition, 30 memory locations per grid point are used for work space
or auxiliary variables.  The code is not optimized for memory usage but
rather for simplicity, so memory savings could be achieved at the
cost of making the code more complicated, less readable and more
time-consuming.  Thus the code requires 86 words of memory per grid
point, or about 24 Mwords for a $64^3$ run.  In practice, the largest
grid  that fit in the 32M queue on the Cray-2 at NCSA was
$70^3$.

The code runs at 140 Mflops on a single processor,
and takes about 0.02 ms of computation time
per grid point and cycle.  Thus a $64^3$ grid runs at about 700 cycles per
hour.  To run the 70 e-foldings of inflation required to obtain the
observable universe from one pre-inflation Hubble volume takes almost 10
hours of Cray-time with this resolution.  The most time is spent on
calculating the connection coefficients $\Gamma^i_{jk}$ and the Ricci
tensor $R_{ij}$, about 43\% of total time.  The second most time is
used  in the evolution of
the extrinsic curvature tensor $A^i_j$, 11\%.

\bigbreak
\centerline{\bf V. INHOMOGENEOUS INFLATION RUNS}
\medskip

We now present some runs of inhomogeneous inflation.  These runs
use the flat initial data described in section II.  A run
using initial data obtained with the initial data solver was presented in
Ref.~\r\LaKuMaK.

The requirement that the initial time slice has homogeneous total energy
density
(as implied by flat initial data) gives
an upper limit for the variation of the inflaton field $\phi$ within a Hubble
length.  From
$$
   {1\over2}(\nabla\phi)^2  <  \rho  = {3\m^2\over8\pi}H^2   \eqno(5.1)
$$
we get
$$
   \nabla\phi < \biggl({3\over4\pi}\biggr)^{1/2}{\m\over H^{-1}}.  \eqno(5.2)
$$
Thus inhomogeneities that begin inside the horizon have small amplitudes.  Runs
where we had large initial variation in the inflaton field have initial grid
lengths $L$ equal to many Hubble lengths.

These runs are of chaotic inflation.  The potential is
$V(\phi) = {1\over4} \lambda\phi^4$, with $\lambda = 10^{-6}$.
We chose initial data of the form
$$
   \phi(t_0,{\bf x}) = \phi_0 + \delta\phi\sum_{l,m,n=1}^2{1\over lmn}
      \sin x_l \sin y_m \sin z_n,  \eqno(5.3)
$$
where $ x_l = 2\pi lx/L + \theta_{xl} $, {\it etc.}, with $\theta_{xl}$
random phases.  The initial Hubble constant was
chosen to be $H = 0.1\m$,  and  $\phi_0 = 5\m$.

We present a small-scale run with $L = H^{-1}$, $\delta\phi = 0.0125\m$,
and a large-scale run with $L = 32H^{-1}$, $\delta\phi = 0.4\m$.
The phases for these particular runs were
$\theta_{x1} = 3.83$, $\theta_{x2} = 2.59$,
$\theta_{y1} = 3.02$, $\theta_{y2} = 1.83$,
$\theta_{z1} = 3.19$, and $\theta_{z2} = 1.33$.
The same phases were used in both runs.  In the initial data, the minimum
$\phi$
value turned out to be $\phi_0-3.3\delta\phi$, and the maximum
$\phi_0+2.1\delta\phi$.  The maximum value of $\nabla\phi$ was within a factor
of two from the upper limit.  In Fig.~6 we show the regions of small and large
$\phi$.

The results are shown as 3-dimensional contour plots of the scalar variables
$\phi$, $\h$, $K$, and $R$.  These are supplemented by 1-dimensional plots
showing the quantities along a reference line at different time slices.  This
reference line
runs parallel to the $y$-axis and goes through the initial $\phi$
minimum.  Figure 6 applies to the initial data of both the small- and the
large-scale runs,
although the {\it amplitude} of these $\phi$ variations were of different
magnitude.

We discuss the small-scale run first.  The grid length was initially set
equal to
one Hubble length.  Thus the inhomogeneities are at first well inside the
horizon.  The dynamical time scale of the inflaton field is therefore
shorter than the expansion time scale, given by the Hubble time.  At the
initial slice the energy density is dominated by the kinetic energy of
the $\phi$ and $\phi_R$ fields, accounting for 84\% of
total energy.  13 \% of the energy
is in $V(\phi)$ and 3\% in field gradients (although at the point of
maximum $\nabla\phi$ they account for more than half of the energy
density).  The initial Hubble time is $t_H = 10\t$.  We arbitrarily set
the initial time as $t_0 = t_H(t_0)/2 = 5\t$.  The time step is at first
controlled by the Courant condition, and thus we get several hundred
time steps per Hubble time.  We show a number of contour plots of the
slice at cycle 256, or $t = 9.2\t$.  Of the total energy in this slice,
52\% is kinetic, 46\% potential, and 2\% gradient.  The Hubble time is
now roughly $18\t$, so this slice is about a quarter of a Hubble time
from the beginning.

 As the inhomogeneities evolve, they oscillate and their amplitude is damped,
see Fig.~7.  After a quarter of a Hubble time of evolution, the regions
of small and large $\phi$ are quite different from what they were initially,
see Fig.~8.  By the time the inhomogeneities have exited the horizon, the
inflaton field has become much more homogeneous than it was initially.

Since we started with flat initial data, $\h$, $K$, and $R$ were all
homogeneous at the initial time (and $R$ was zero).  The inhomogeneities in
$\phi$  induce inhomogeneities in these curvature quantities.

We show regions
of small and large expansion $\h$ after a quarter of a Hubble time in Fig.~9.
Regions of both large and small initial $\phi$ end up as regions of small
expansion.  The region of largest expansion forms a shell surrounding
the initial $\phi$ minimum.  This is partly due to the large gradient
energies in this region at early times, and partly due to the inflaton
field happening to have large potential values in this region during the time
when most of the expansion inhomogeneity is generated.
After a volume expansion of about 100, further expansion is rather
homogeneous, and the $\h$ profile stays the same, see Fig.~10.

The extrinsic curvature $K$ is related to the rate of change in $\h$.
In Fig.~11 large absolute values of $K$ in the
region of large $\h$ are seen.  Later $K$ becomes rather homogeneous (see
Fig.~12).  Regions of largest intrinsic curvature $R$ (Fig.~13) seem to
be at or around regions of small expansion.  This is partly because in
this run those are more pronounced than regions of large expansion,
partly because expansion decreases $R$ by increasing the curvature
radius.
Figure 14 shows $R$ along our reference line at different times.

We show only the early part of this run, since the later
evolution is less interesting.  The inflaton rolls down its
potential staying rather homogeneous.  Inflation ends as $\phi$ reaches
the bottom and begins to oscillate.

In the large-scale run the inhomogeneities are outside the horizon.
They do not oscillate.  Rather, they maintain their shape without
damping.  After a few Hubble times the regions of low (Fig.~15) and high
$\phi$ look very much the same as initially (Fig.~6).  Figure 16 shows
$\phi$ along the reference line at various times through the entire run
past the end of inflation.  We see that aside from the flattening of the
sharp initial minimum, the inhomogeneity maintains its shape.
Because of the larger scale, the Courant condition allows a larger
initial time step for this run.

The spatial pattern of the other scalar quantities follows that of
$\phi$.  Larger values of $\phi$ lead to faster expansion.  At first the
effect of the other scalar field $\phi_R$ (see end of section II) shows,
but later the regions of the most expansion match those of large $\phi$
rather closely (Fig.~17).  Since the regions of large $\phi$ and high
inflation stay the same, the inhomogeneity in the expansion becomes
steeper and steeper (Fig.~18).  When inflation ends, some regions have
expanded by a volume factor of more than $10^{130}$, while others have
expanded by less than $10^{75}$.

For the extrinsic curvature we show just the 1-d plot (Fig.~19).  The
3-d plots would look intermediate between the $\phi$ and $\h$ plots.
Note the correspondence between Figs.~16, 18, and 19.  Larger $\phi$,
and thus $V(\phi)$, means faster expansion, {\it i.e.}, larger (more
negative) $K$.  This then builds up as a larger cumulative expansion
$\h$.

In Fig.~20 we show regions of large positive and negative 3-curvature
$R$ at cycle 64, after half a Hubble time of evolution.  The largest
curvature is at the region where the expansion has been the smallest.
This region of large positive curvature is surrounded by a shell of
negative curvature.  After more inflation, the curvature decreases but
the relative contrast between large- and small-curvature regions
increases, so that later contour plots would pick up only the
largest-curvature region near the back corner.

The Hamiltonian and momentum constraints were monitored during the
evolution.  During the small-scale run the maximum errors
(root-mean-square relative difference between the left and right
sides) were:  Hamiltonian 0.2\%, momentum 3.3\%.  In the large-scale run
these errors were smaller:  Hamiltonian 0.05\%, momentum 0.3\%.  We ran these
runs also with a coarser grid.  Going from a $32^3$ grid to the $64^3$
grid, the Hamiltonian error went down to about one-quarter, the momentum
error to about one-half.

\bigbreak
\centerline{\bf VI. CONCLUSIONS}
\medskip

We have created a working truly 3-dimensional numerical relativity code
for studying inhomogeneous inflation.  The code has been tested against
perturbation theory with good results.  On Cray Y-MP computers,
a practical grid size to run the code with is around $48^3$ to $64^3$.
This is enough to handle an interesting amount of
structure with reasonable accuracy (but just barely).  Present memory
limits would allow larger grids, but the increased computation
time leads to much worse throughput.

A variant of this code is being used\r\LaKuMaP\ at present for interacting
black hole studies, with initial data computed as described by
Cook {\it et al.},\r\Cook\ and with asymptotically flat (rather than periodic
or
reflective) boundary conditions.  Sizes at least up to $128^3$ are being
implemented on a CM-5 computer\r\Smarr).  We are
working also to move
the cosmology code to this machine where it will be run with inflaton and
with other sources, to describe a number of 3-dimensional physical situations.

One of the main motivations for the inflationary scenario was to explain
the large-scale homogeneity and isotropy of the universe without
requiring these properties as initial conditions\r\GolPirR.
An apparent shortcoming in the main body of work on inflation is then,
that it has nevertheless been in the context of homogeneous space or
small perturbations around it.  We have now presented numerical
simulations of inhomogeneous inflation, where the inhomogeneity has been
truly 3-dimensional and nonperturbative.  These
runs had sufficient inflation to solve the standard cosmological
conundra.  In particular, they contain regions where evidence of the
initial inhomogeneity would not be locally observable.  We have thus
demonstrated the viability of inflation with inhomogeneous initial
conditions.

\bigbreak
\centerline{\bf ACKNOWLEDGEMENTS}
\medskip

Computations were carried out at the National Center for Supercomputing
Applications, University of Illinois at Urbana--Champaign.
Work performed in part under the auspices of the U.~S.~Department of Energy
by the Lawrence Livermore National Laboratory under contract
W--7405--ENG--48, and supported in part by NSF PHY8806567.
Work by Richard Matzner supported also by a Cray
Research University grant, and by Texas Advanced Research Program grant
ARP-085.

\bigbreak
\centerline{\bf References}
\medskip

\def\i#1{\par\hang\indent\llap{[#1]\quad}\ignorespaces}
{\frenchspacing
\i\Guth    A. Guth, Phys. Rev. D {\bf 23}, 347 (1981).
\i\Olive   K. A. Olive, Phys. Rep. {\bf 190}, 307 (1990).
\i\Smoot   G. F. Smoot {\it et al.}, Astrophys. J. {\bf 396}, L1 (1992).
\i\KCMW    H. Kurki-Suonio, J. Centrella, R. A. Matzner, and J. R.
           Wilson, Phys. Rev. D {\bf 35}, 435 (1987).
\i\HPV     K. A. Holcomb, S. J. Park, and E. T. Vishniac, Phys. Rev. D
           {\bf 39}, 1058 (1989)
\i\GolPir  D. Goldwirth and T. Piran, Phys. Rev. D {\bf 40}, 3263 (1989).
\i\GolPirK D. Goldwirth and T. Piran, Phys. Rev. Lett. {\bf 64}, 2852 (1990).
\i\LaKuMaI P. Laguna, H. Kurki-Suonio, and R.A. Matzner, Phys. Rev. D {\bf 44},
           3077 (1991).
\i\MTW     C.W. Misner, K.S. Thorne, and J.A. Wheeler, {\it Gravitation}
           (Freeman, San Francisco, 1973).
\i\Linde   A. D. Linde, Phys. Lett. B {\bf 129}, 177 (1983).
\i\Bra     R. Brandenberger, R. Kahn, and W. H. Press, Phys. Rev. D {\bf 28},
           1809 (1983); R. Brandenberger and R. Kahn, Phys. Rev. D {\bf 29},
           2172 (1984); R. H. Brandenberger, Rev. Mod. Phys. {\bf 57}, 1
(1985).
\i\LaKuMaK P. Laguna, H. Kurki-Suonio, R.A. Matzner, Proceedings of the Sixth
           Marcel Grossmann Meeting on General Relativity, (H. Sato and T.
           Nakamura, Eds.), p. 1343, World Scientific (Singapore 1992).
\i\LaKuMaP P. Laguna, H. Kurki-Suonio, R.A. Matzner (in progress, 1993)
\i\Cook    G.R. Cook, M.W. Choptuik, M.R. Dubal, S. Klasky, R.A. Matzner, and
           S.R. Oliveira, Phys. Rev. D {\bf 47}, 1471 (1993).
\i\Smarr   L. Smarr, private communication (1993).
\i\GolPirR D. S. Goldwirth and T. Piran, Phys. Rep. {\bf 214}, 223 (1992).
}

\vfil\eject


\bigskip
\centerline{\bf Figure captions}
\medskip

FIG.~1.  The inflaton field $\phi$ as a function of time in homogeneous
chaotic inflation.  The lines showing the code result and the exact
result fall on top of each other and cannot be distinguished by eye.
Inflation ends near $t = 300\t$.

FIG.~2.  Same as Fig.~1, but for the scale factor $a$.  Inflation
ends after a linear expansion of $10^{34}$, or a volume expansion of
$10^{102}$.

FIG.~3.  The amplitudes of the three perpendicular plane-wave
perturbations $\delta\phi_1$, $\delta\phi_2$, and $\delta\phi_3$ in the
inflaton field.  The different line styles show code results with
different grid resolutions, as well as the perturbation theory result.
(a)  Early part of the run, perturbations inside the horizon:  The
perturbations
with the longer wavelengths have longer oscillation periods, but all are
damped by the expansion at the same rate.  In this case the code results
are rather accurate, and the lines fall onto each other.
(b)  Late part of the run, perturbations outside the horizon: The
vertical scale has been expanded by a factor of 25.  The inset displays
the last decade (from $t = 10^3\t$ to $t = 10^4\t$) with a further
expanded scale to show the post-inflation oscillations.

FIG.~4.  Same as Fig.~3, but for the perturbations $A_1$, $A_2$, and
$A_3$ in the metric.  (a) Early part of the run.  (b)  The entire run.
The perturbations exit the horizon between $t = 2\t$ and $t = 3\t$ and
the oscillatory behaviour changes to monotonic growth, which ceases as
inflation ends.

FIG.~5.  Same as Fig.~3, but for the perturbations $B_1$, $B_2$, and
$B_3$ in the metric.  We only show the early part of the run, because in
the late part $B$ stays almost constant in time.

FIG.~6.  Regions of small and large values of $\phi$ in the initial
data.  We express the contour levels as a percentage between the minimum
(0\%) and the maximum (100\%) value in the slice.  (a)  The regions of
small $\phi$ are enclosed by 20\%, 40\% and 50\% contours.  In the
small-scale run these correspond to $\phi = 4.973\m$, $\phi = 4.986\m$,
and $\phi = 4.993\m$.  In the large-scale run the contours have $\phi =
4.12\m$, $\phi = 4.55\m$, and $\phi = 4.77\m$.  The minimum value, near the
back corner enclosed by all three contours, was $4.959\m$ in the
small-scale run, and $3.69\m$ in the large-scale run.
(b) The regions of large $\phi$ are shown enclosed by 75\% and 90\%
contours ($5.010\m$ and $5.020\m$ for the small-scale run,
$5.31\m$ and $5.63\m$ for the large-scale run).  The maximum value was
$5.026\m$ in the small-scale run, and $5.85\m$ in the large-scale run.
In some of the following figures, quantities are plotted along a
reference line. This line runs parallel to the $y$-axis, which in this
figure goes to the right from the origin.  The reference line
goes through the initial
$\phi$ minimum near the back corner and through the  region with large
$\phi$ near the right end of the $y$-axis.

FIG.~7.  Figures 7--14 show results from the small-scale run.
This figure shows the inflaton field along the reference line at various time
slices.  These correspond to $t/\t = 5$ (initial time), 5.6, 6.4, 7.6,
9.2, 11.3, 14.3, 18.5, 24.1, 42.6, and 77.7.  The Hubble time was
initially $10\t$.  At the last slice shown, where inflation has taken
hold, the potential contributing 99.6\% of total energy, the approximate
Hubble time is $28\t$.

FIG.~8.  The inflaton field at $t = 9.2\t$.  We show regions of
small $\phi$ enclosed by 15\% and 30\% contours.

FIG.~9.  The local volume expansion $\h$ at $t = 9.2\t$.  (a) Regions of
small $\h$ are enclosed by 30\%, 70\%, and 83\% contours.  (b) Regions of
large $\h$ are enclosed by 94\% and 98\% contours.

FIG.~10.  Same as Fig.~7 but for $\h$.

FIG.~11.  Trace of the extrinsic curvature at $t = 9.2\t$.  (a) Small
values of $|K|$ are enclosed by 50\%, 80\%, and 88\% contours.
(b) Large values of $|K|$ are enclosed by the 98\% contour.

FIG.~12.  Same as Fig.~7 but for $K$.

FIG.~13.  Scalar curvature at $t = 9.2\t$.  Regions of large $R$ are
enclosed by 25\% and 75\% contours.

FIG.~14.  Same as Fig.~7 but for $R$.

FIG.~15.  Figures 15--20 show results from the large-scale run.  This
figure shows the inflaton field at cycle 512, or $t = 166\t$, after
about 6 Hubble times.  Regions of small $\phi$ are enclosed by 15\%,
34\%, and 44\% contours.

FIG.~16.  The inflaton field along the reference line at every 512th
cycle.  The initial slice is at the top, and the last slices at the bottom,
as $\phi$ moves down with time.  The slices shown correspond to
$t/\t = 5$, 166, 482, 598, 852, 1140, 1471, 1861, 2337, 2947, 3797,
5201, 9624, $6.02\times10^4$, and $7.10\times10^5$.  Defining the end of
inflation as the time when $V(\phi)$ falls below 50\% of the total energy,
it happens at $t = 1.2\times10^4\t$.  The run contains 4 periods of
post-inflation oscillations.  The last two slices shown (close to each
other) are from this era.

FIG.~17.  The local volume expansion $\h$. (a) At cycle 64 or $t = 14.7\t$.
Regions of large $\h$ are enclosed by 60\% and 80\% contours. (b) At
cycle 512 or $t = 166\t$.  Regions of large $\h$ are enclosed by 5\% and
50\% contours.

FIG.~18.  Same as Fig.~16 but for $\h$.  $\h$ moves up with time.

FIG.~19.  Same as Fig.~16 but for $K$. $K$ moves up with time.
The initial slice, where $K$ is homogeneous, falls below the area plotted.

FIG.~20.  Scalar curvature at $t = 14.7\t$.  (a) Regions of large
positive $R$ enclosed by 15\%, 30\%, and 50\% contours
$(R = 0.15\times10^{-3}\m^2, 0.55\times10^{-3}\m^2,
1.10\times10^{-3}\m^2)$.
(b) Regions of negative $R$ enclosed by the 3.7\% contour
$(R = -0.16\times10^{-3})$.

\bye